\documentclass[journal]{IEEEtran} 

\usepackage{cite}
\usepackage[dvips]{graphicx}

\usepackage[cmex10]{amsmath}
\usepackage{amssymb}
\usepackage{algorithmic}
\usepackage{algorithm}
\usepackage{amsfonts}
\usepackage{array}
\usepackage{mdwmath}
\usepackage{mdwtab}
\usepackage{stfloats}
\usepackage{color}

\usepackage{soul,color}

\makeatletter

\newcommand{\Rmnum}[1]{\expandafter\@slowromancap\romannumeral #1@}
\makeatother

\begin{document}
%
\title{Channel Modeling and Performance Analysis of Airplane-Satellite Terahertz Band Communications}
%
%
%


\author{
\IEEEauthorblockN{Joonas~Kokkoniemi$^\star$,~\IEEEmembership{Member,~IEEE,}
~Josep~M.~Jornet, ~\IEEEmembership{Member,~IEEE,}~Vitaly~Petrov, ~\IEEEmembership{Student~Member,~IEEE,}\\
~Yevgeni~Koucheryavy, ~\IEEEmembership{Senior~Member,~IEEE,}~
and~Markku~Juntti ~\IEEEmembership{Fellow,~IEEE}}

\thanks{$^\star$Corresponding author. Email: joonas.kokkoniemi@oulu.fi.}
\thanks{Joonas Kokkoniemi and Markku Juntti are with the Centre for Wireless Communications, University of Oulu, Oulu, Finland.}
\thanks{Josep M. Jornet is with the Department of Electrical and Computer Engineering, Northeastern University, Boston, MA, USA.}
\thanks{Vitaly Petrov and Yevgeni Koucheryavy are with the Laboratory of Electronics and Communications Engineering, Tampere University, Tampere, Finland.}
\thanks{This work was supported by Horizon 2020, European Union's Framework Programme for Research and Innovation, under grant agreement no. 761794 (TERRANOVA) and no. 871464 (ARIADNE). It was also supported in part by the Academy of Finland 6Genesis Flagship under grant no.~318927.}}

\maketitle

\begin{abstract}
Wireless connectivity in airplanes is becoming more important, demanded, and common. One of the largest bottlenecks with the in-flight Internet is that the airplane is far away from both the satellites and the ground base stations during most of the flight time. Maintaining a reliable and high-rate wireless connection with the airplane over such a long-range link thus becomes a challenge. Microwave frequencies allow for long link distances but lack the data rate to serve up to several hundreds of potential onboard customers. Higher bands in the millimeter-wave spectrum (30\,GHz--300\,GHz) have, therefore, been utilized to overcome the bandwidth limitations. Still, the per-user throughput with state-of-the-art millimeter-wave systems is an order of magnitude lower than the one available with terrestrial wireless networks. In this paper, we take a step further and study the channel characteristics for the terahertz band (THz, 0.3\,THz--10\,THz) in order to map the feasibility of this band for aviation. We first propose a detailed channel model for aerial THz communications taking into account both the non-flat Earth geometry and the main features of the frequency-selective THz channel. We then apply this model to estimate the characteristics of aerial THz links in different conditions. We finally determine the altitudes where the use of airplane-to-satellite THz connection becomes preferable over the airplane-to-ground THz link. Our results reveal that the capacity of the airborne THz link may reach up to 120 Gbit/s, thus enabling cellular-equivalent data rates to the passengers and staff during the entire flight.

\end{abstract}


%
\IEEEpeerreviewmaketitle

\section{Introduction}
\label{section:intro}

\IEEEPARstart{T}{oday}, terahertz (THz, 0.3\,THz--3\,THz) communication is widely considered to be one of the next frontiers for future wireless systems. Data transmission over the THz band offers at least an order of magnitude higher rates than emerging millimeter-wave (mmWave, 30\,GHz--300\,GHz) communication systems and two orders of magnitude when compared to state-of-the-art microwave solutions~\cite{thz_magazine1,thz_magazine2}. Hence, THz communication cannot only exceed the performance requirements for the fifth-generation (5G) networks, but also enable many tempting applications in the area of holographic communications, augmented and virtual reality, and tactile internet~\cite{sundeep_6g_commag}. While the first prototypes of point-to-point connections over the low THz frequencies are already appearing~\cite{rap_6g_survey}, the research community is slowly switching its focus on studying the features of THz links in prospective application-specific setups.

One of the attractive usage scenarios in the area is enabling THz connectivity with flying airplanes. The aviation industry has been growing by over 6\%~a year over the recent decades, reaching an incredible figure of 4.3~billion passengers carried in 2018~\cite{icao_report}. Although the current COVID-19 pandemic is momentarily slowing down the development, in the long run the trend is expected to continue. While the average duration of a flight is less than 3~hours, a substantial amount of flights are longer than 7~hours. Some flights between the continents take the even longer time up to enormous 18~hours 45~minutes with a recently introduced route between Singapore and Newark, USA, by Singapore Airlines. Despite the continuously increased popularity of long haul flights, the airplane is one of the few typical locations where the average person is used to not have high-rate Internet connectivity.

Aiming to address this limitation, modern aircrafts are already equipped with satellite-based solutions operating in Ku (12--18~GHz) and, recently, Ka (26.5--40~GHz) frequency bands. Still, these systems can provide only limited service to restricted numbers of passengers, as the aggregated traffic from over 400~users (e.g., in a Airbus A350-1000) cannot be efficiently multiplexed into a single mmWave link. Here, the use broadband THz spectrum provides decisive advantages, rapidly increasing the data rate of the airplane connectivity and thus allowing all the passengers and crew members to stay continuously connected to their common applications and services. However, the high-rate airborne connectivity in the THz band requires not only the evolution in the transceiver and antenna design. Importantly, it also demands a better understanding of the properties of the THz signal when propagating through the atmosphere at a high altitude. Potentially high losses at a multi-kilometer THz link between the airplane and a satellite may even question the feasibility of the airborne THz communications.

\subsection*{Related Work}
A general model for free-space wireless communications in the THz band was first presented in~\cite{Jornet2011}. The model enables estimating the path loss over a direct THz link, properly accounting for the specific effects present in THz communications, such as molecular absorption. The properties of the THz propagation in complex environments have been extensively studied via ray-based simulations~\cite{kurner1,petrov_commag_thz,kurner2,petrov_wcm_radar}. For this purpose, deterministic channel models are first built following the exact geometry of the scenario. Then, the impact of multipath propagation is modeled following either \emph{ray-tracing} or \emph{ray-launching} approach. The approach is featured by high accuracy but limited scalability, as even the minor modifications in the modeled scenario requires a restart of the computationally complex ray-based simulations. In response to this issue, mathematical approaches to model multi-path THz communications have been proposed in~\cite{chong_multipath} and~\cite{kurner_stochastic_channel_models}. These stochastic models primarily target indoor THz systems, where the signal can reflect from walls and other obstacles multiple times before reaching the target receiver.

Despite the progress in understanding the key features of the THz signal propagation, the airborne nature of the THz links between an airplane and a satellite has its own specifics that must be taken into account in channel modeling. The topic has been partially covered in prior works. A simulation-based model for THz band satellite links has been proposed in~\cite{modeling_trans_thz_science_2015}. Later, a review of weather impact on outdoor THz links has been presented in~\cite{federici_nanocomnet_2016}. The latter study has been complemented by Y.~Balal and Y.~Pinhasi in~\cite{balal_atmospheric_2018}, where effects of the atmosphere inhomogeneous refractivity on mmWave and THz band satellite links has been explored.

In parallel to academic research, some initial steps towards characterizing the airborne THz links have been made by International Telecommunication Union (ITU). Specifically, the approach to estimate the signal attenuation by atmospheric gases has been presented in ITU-R P.676-9~\cite{ITU_R_P_676_9} while the noise levels for different frequencies (including the THz band) have been estimated in ITU-R P.372-13~\cite{ITU_R_P_372_13}.

Finally, one of the closest models to our approach -- the am atmospheric model -- has been recently presented by S. Paine from Smithsonian Astrophysical Observatory in~\cite{Paine2012}. The model first characterizes the absorption losses at different altitudes and then allows the estimation of the average effect when propagating through several atmospheric layers by integrating the obtained data. However, the curvature of the atmosphere has not been fully taken into account, which, as will be shown later in the article, appears to notably affect the accuracy of the model in our target plane-to-satellite scenario.

\subsection*{Our Motivation and Contribution}

Summarizing the related work survey, no model has been proposed to date for airborne THz band communications that takes into account: (i) the curvature of the atmosphere and, consequently, (ii) the non-uniformity of atmospheric absorption losses. Simultaneously, there has not yet been presented a detailed numerical study investigating the feasibility of airborne wireless communications via the THz band while accounting for the above-mentioned specifics. We aim to partially fill the gap in this article.

The main contributions of this article are thus:
\begin{itemize}
\item The mathematical model to characterize the airborne THz band communications is proposed, taking into account (i) the features of the target use case and the deployment geometry, (ii) the peculiarities of the signal propagation through the atmosphere, as well as (iii) the prospective characteristics of THz radio equipment to operate in the airborne scenarios.

\item The illustrative numerical study is performed estimating the performance of the data exchange between an airplane and a satellite over the THz frequencies. Particularly, the signal-to-noise ratio (SNR) and the mean capacity of the airborne THz link are estimated. Our numerical study allows to conclude that up to $120$\,GBits/s airborne THz links can be maintained using certain sub-bands in the THz spectrum.
\end{itemize}

The remainder of this paper is organized as follows. In Sec.~\ref{section:background} we detail the system model used in our study. The specifics of the THz signal propagation are discussed in detail in Sec.~\ref{sec:thz_propagation}. The channel model for airborne wireless communications in the THz band is introduced in Sec.~\ref{sec:general_channel_model}. We later apply the contributed model in Sec.~\ref{sec:snr} for the received signal quality and performance evaluation of prospective airborne THz links. The obtained analytical results are numerically elaborated in Sec.~\ref{section:res}. The concluding remarks are drawn in the last section.

\section{System Model}
\label{section:background}
\label{sec:system_model}

\subsection{Scenario Description}

\begin{figure}[t!]
    \centering
    \includegraphics[width=3.2in]{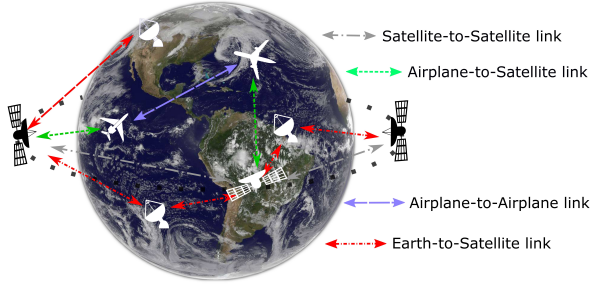}
    \caption{An illustration of the system model considered in this paper and the several wireless links in it. Picture of Earth by NASA/NOAA's GOES Project \cite{NASApic}.}
    \label{fig:sys}
\end{figure}

The general system concept is to provide high data rate fronthaul connection to an airplane, which furthermore serves the passengers on board and in general provides Internet connection to the airplane. However, as a side product, the channel models derived in this paper cover several use cases for aerial vehicle communications, as illustrated in Fig.~\ref{fig:sys}. The considered application scenarios for the channel models in this paper include: (i) airplane-to-satellite (A2S) or satellite-to-airplane (S2A), (ii) Earth-to-airplane (E2A) or airplane-to-Earth (A2E), and (iii) Earth-to-Satellite (E2S) or satellite-to-Earth (S2E). The satellite-to-satellite (S2S) case is not considered in this paper due to strong focus on modeling the molecular absorption and the atmosphere in general. Even the low Earth orbit (LEO) satellites lie in nearly empty space where any atmospheric effects will be small. Therefore, the channel mainly comprises free space loss and the antenna systems.

The target scenario for the A2S and S2A channel modeling is shown in Fig.~\ref{fig:sysgeom1}. The target scenario for channel modeling consists of an airplane at an altitude $h_a$ and a satellite at an altitude $h_{s}$ from the Earth and distance $r_{as}$ from the airplane. The boundary of the atmosphere is $r_\text{atm}$ away from the airplane. As the separation distance between the considered airplane and satellite is significant, the relative speed of the airplane is considered negligible for the channel modeling. The most important parameters and variables are given in Table~\ref{table:params}.


\subsection{Atmospheric Propagation in the Terahertz Band}
\label{sec:thz_propagation}

The main phenomena affecting the propagation of THz signals in the target scenarios are went through in the following. Those include free space path loss and the molecular absorption loss. Some additional loss mechanism include rain, cloud, fog, and possible scattering losses that are briefly discussed below.

\subsubsection{Spreading Loss}
The spreading loss measures the fraction of the power radiated by an isotropic transmitter at a frequency $f$ that an isotropic receiver at a distance $r$ can detect. Under the assumption of spherical propagation, the spreading loss is given by
\begin{equation}
\label{equation:FSPL}
    \left|H_{spr}\left(f,r\right)\right|^2=\frac{1}{4\pi r^2} \frac{\left(c/f\right)^2}{4 \pi}=\left(\frac{c^2}{4\pi f r}\right),
\end{equation}
where $c$ is the speed of light in the medium. In our system, directional antennas (e.g., horn lens antennas, Cassegrain parabolic antennas or beamforming antenna arrays) will expectedly be utilized at the transmitter and the receiver to counter the high channel losses. In that case, their directivity gain $D\left(\theta,\gamma\right)$ with respect to an ideal isotropic emitter and detector (commonly given in dBi) needs to be accounted for.

\subsubsection{Molecular Absorption Loss}
The molecular absorption loss measures the fraction of electromagnetic energy that is converted into kinetic energy internal to vibrating molecules~\cite{Paine2012,SpecCalc,Jornet2011}. For an homogeneous medium with thickness $r$, this is given by the Beer-Lambert law and can be written as:
\begin{equation}
\label{equation:tau_tot}
\left|H_{abs}\left(f,r\right)\right|^2 = e^{-\Sigma_{i}\kappa_{a}^i(f)r},
\end{equation}
where $f$ is the frequency, $r$ is the distance from Tx to Rx and $\kappa_{a}^i(f)$ is the absorption coefficient of $i$th absorbing species (molecule or its isotopologue) at frequency $f$. We denote the total summed absorption coefficient with $\kappa_{a}(f)$, i.e., $\kappa_{a}(f) \equiv \sum_i\kappa_{a}^i(f)$.

In our system, the medium is not homogeneous, as the density of different absorbing molecules (particularly water vapor molecules) changes drastically with altitude. In the case of vertical paths, the absorption coefficient $\kappa_a$ depends on distance $\kappa_a\left(f,r\right)$ and its variations need to be accurately computed along the signal propagation path.

\subsubsection{Scattering Loss on Aerosols}

The one additional loss mechanism comes from the aerosol scattering. The aerosols are small particles suspended by air, such as dust, ice particles, pollution, etc. Those are modelled by the Beer-Lambert law and therefore can theoretically cause significant losses over long distance links \cite{Kokkoniemi2015}. However, as it was shown in \cite{Kokkoniemi2015}, the THz frequencies require rather large particles to cause significant losses. On Earth this is possible, e.g., in very dusty conditions, but in general this is not a problem at higher altitudes. The possible losses on clouds can be handled as shown below. For the sake of tractability, we ignore these losses in the further study.

\subsubsection{Rain and Cloud losses}
Additional attenuation may be caused by adverse weather conditions in A2A, A2S, and S2A scenarios. The ITU-R provides rain \cite{ITU838} and cloud/fog \cite{ITU840} attenuation models. It should be noted that the cloud attenuation is formally valid only up to 200 GHz frequencies \cite{ITU840}. Furthermore, the cloud densities, cloud thicknesses and rain rates vary greatly depending on the type of the cloud, and the probability of different types of clouds vary geographically. However, in any case the Earth to airplane or satellite paths experience added path losses ranging from few decibels up to tens of decibels in the presence of clouds and depending on the weather in general.

In the further study, we define the rain and cloud losses with the distance-dependant variables $\delta_\text{Rain}(r_r)$ and $\delta_\text{Cloud}(r_c)$, respectively.


\section{Atmospheric Path Loss Modeling}
\label{sec:general_channel_model}

In this section, we derive the path loss for the above aerial use cases. As highlighted in the previous section, the main challenge arises from the non-uniform molecular absorption coefficient through the atmosphere. The density, temperature, and molecular composition of the atmosphere depend on the altitude. This section gives the dynamic molecular absorption loss coefficient valid at any altitude by taking into account the impact of atmospheric parameters on the absorption line calculations. After that, we go through the transmission path geometry of slant paths through the atmosphere. Table \ref{table:params} gives the most important constants, variables, and parameters utilized in this paper.

\begin{table}[t!] 
\caption{The most important constants, variables, and parameters used in the paper.}
\centering
\begin{tabular}{ccc} 
\hline\hline
\multicolumn{3}{c}{\bf Constants} \\
\hline
\bf Constant & \bf Symbol & \bf Value \\
\hline
Radius of Earth & $R$ & 6,371,000 m\\
Speed of light & $c$ & 299,792,458 m/s\\
Standard pressure & $p_0$ & 101325 Pa\\
Standard temperature & $T_0$ & 296 K\\
\hline\hline
\multicolumn{3}{c}{\bf Variables} \\
\hline
\multicolumn{2}{c}{\bf Variable} & \bf Symbol \\
\hline
\multicolumn{2}{c}{Absorption coefficient [1/m]} & $\kappa_a(f)$\\
\multicolumn{2}{c}{Center frequency [GHz]} & $f_c$\\
\hline\hline
\multicolumn{3}{c}{\bf Parameters} \\
\hline
\multicolumn{2}{c}{\bf Parameter} & \bf Symbol \\
\hline
\multicolumn{2}{c}{Altitude of airplane [m]} & $h_a$\\
\multicolumn{2}{c}{Altitude of satellite [m]} & $h_s$\\
\multicolumn{2}{c}{Angle to zenith [rad]} & $\theta$\\
\multicolumn{2}{c}{Depth of the atmosphere [m]} & $r_\text{atm}$\\
\multicolumn{2}{c}{Distance from plane to satellite [m]} & $r_{as}$\\
\multicolumn{2}{c}{Frequency [Hz]} & $f$\\
\multicolumn{2}{c}{Latitude of airplane [rad]} & $\phi_a$\\
\multicolumn{2}{c}{Latitude of satellite [rad]} & $\phi_s$\\
\multicolumn{2}{c}{Pressure [Pa]} & $p$\\
\multicolumn{2}{c}{Temperature [K]} & $T$\\
\hline\hline
\end{tabular}
\label{table:params}
\end{table}

\subsection{Non-homogeneous Molecular Absorption Loss}
\label{section:ch}
Traditionally the channel modeling for the THz frequencies has been considered for terrestrial communications. When moving to higher altitudes, the homogeneous assumption for the line shape functions is no longer valid. We can also utilize meteorological data to take into account the global variations in water vapor content as a function of latitude, longitude, and altitude. Those will be discussed briefly in conjunction with the numerical results. Similarly, when moving to higher altitudes, mixing ratios of all the molecular species change and has to be taken into account for accurate molecular absorption modeling.




The absorption loss was described above and it depends on the link distance and absorption coefficient. The latter depends on pressure, temperature and molecular composition as
\begin{equation}
\kappa_{a}^i(f) = N_i\sigma_i(f),
\label{equation:kappaTot}
\end{equation}
where $N_i$ is the number density and $\sigma_i(f)$ is the absorption cross section of the $i$th absorbing species. The number density $N_i$ for the absorbing species $i$ can be estimated by the ideal gas law \cite{Paine2012,Jornet2011}
\begin{equation}
N_i = \frac{p}{p_0}\frac{T_0}{T}\mu_in_0 = \frac{p}{p_0}\frac{T_0}{T}\frac{p_0}{RT_0}\mu_iN_A = \frac{p}{RT}\mu_iN_A,
\label{equation:Q}
\end{equation}
where $p_0$ and $T_0$ are the standard pressure and temperature (101325 Pa and 296 K, respectively), $p$ is the pressure, $T$ is the temperature, $\mu_i$ is the volume mixing ratio of absorbing species $i$, $R$ is gas constant ($R = k_BN_A$), $k_B$ is the Boltzmann constant, $N_A$ is the Avogadro constant and $n_0 = p_0 N_A/(RT_0)$ the number density of the molecules in standard pressure and temperature.

The different molecules have specific abundances in atmosphere. In addition, the different isotopologues of the molecules have their natural abundances \cite{Paine2012,Jornet2011}. We have included all the abundances of molecules and their subspecies into the variable $\mu_i$. The abundances of atmospheric molecules and the other parameters for line-by-line calculations can be found in high-resolution transmission molecular absorption database (HITRAN) \cite{HITRAN08}. Other similar models to HITRAN, such as GEISA \cite{Jacquinet-Husson2011} and JPL \cite{Pickett2003}, also exist.


The absorption cross section $\sigma_i(f)$ is calculated as a product of spectral line intensity $S^i(T)$ and spectral line shape $F^i(p,f,T)$ ($\sigma_i(f) = S^i(T)F^i(p,f,T)$). The absorption cross section gives the effective absorption area for a single particle. The spectral line intensity gives the absorption strength of the spectral lines and spectral line shape gives the shape of the absorption lines. The line shape is normalized so that the integration over the line shape equals unity \cite{Pierrehumbert2011}. It should be noticed that line intensity depends only on temperature and line shape depends on frequency, pressure and temperature. For simplicity, we will use notation $F^i(f)$ for line shape, as the line shape is calculated for each line center $f^i_0$. In order to calculate the line shape $F^i(f)$, the resonance frequencies $f_c^i$ of the line centers $f_0^i$ must be calculated first. The resonance frequencies increase from zero-pressure position $f_0^i$ according to \cite{SpecCalc,Jornet2011}
\begin{equation}
f^i_c = f^i_0 + \delta_i\frac{p}{p_0},
\end{equation}
where $\delta_i$ is the linear pressure shift.

Even though the absorption process is discrete in frequency domain, the individual absorption lines are spread due to collisions between the molecules (Lorentz broadening), as well as because of the velocity of the molecules (Doppler broadening). The pressure broadening can be expressed with Lorentz half-width $\alpha^i_L$ (at pressures higher than $10$ kPa \cite{SpecCalc}). This can be obtained from foreign and self-broadened half-widths $\alpha^{f}_0$ and $\alpha^i_0$ respectively by \cite{Paine2012,SpecCalc}
\begin{equation}
\alpha^i_L = [(1 - \mu_i)\alpha^{f}_0 + \mu_i\alpha^i_0]\left(\frac{p}{p_0}\right)\left(\frac{T_0}{T}\right)^\gamma,
\end{equation}
where $\gamma$ is temperature broadening coefficient. Self-broadening is caused by the collisions between molecules of the same species, while foreign-broadening is due to the inter-molecular collisions. The coefficients $\gamma$, $\alpha^{f}_0$ and $\alpha^i_0$ can be obtained from the line catalogues.

At high air pressure, Lorentzian line shapes are utilized \cite{Huang2004}. The most familiar of those is the Lorentz line shape \cite{Paine2012,VanVleck1945}
\begin{equation}
F^i_{L}(f\pm f^i_c) = \frac{1}{\pi}\frac{\alpha^i_L}{(f \pm f^i_c)^2+(\alpha^i_L)^2}.
\end{equation}
This was enhanced by Van Vleck and Weisskopf in 1945 \cite{VanVleck1945} and the Van Vleck-Weisskopf asymmetric line shape is defined as \cite{VanVleck1945,VanVleck1977,Halevy2009}
\begin{equation}
F^i_{VVW}(f) = \left(\frac{f}{f^i_c}\right)^2[F^i_{L}(f-f^i_c)+F^i_{L}(f+f^i_c)].
\end{equation}
The Van Vleck-Weisskopf line shape with far end adjustments can be obtained as in \cite{SpecCalc}
\begin{equation}
F^i_{VVH}(f) = \frac{f}{f^i_{c}}\frac{\tanh(\frac{hf}{2k_BT})}{\tanh(\frac{hf^i_c}{2k_BT})}[F^i_{L}(f-f^i_c)+F^i_{L}(f+f^i_c)],
\end{equation}
where $h$ is Planck constant. This line shape has also been referred to as Van Vleck-Huber line shape due to derivation by Van Vleck and Huber \cite{VanVleck1977}. In reality, the differences between these line shapes are rather small in the THz band and the choice of one over another does not produce large error.

At low pressures (below one kPa), i.e., in the higher altitudes, the Doppler broadening becomes the most significant broadening mechanism and it causes a Gaussian line shape \cite{SpecCalc}. The Doppler broadening half-width can be expressed as
\begin{equation}
\alpha_D^i = \frac{f_c^i}{c}\sqrt{\frac{2\log(2)k_BT}{m_ic^2}},
\end{equation}
where $m_i$ is the molar mass of the absorbing species $i$. The Doppler line shape can be obtained as
\begin{equation}
F^i_{D}(f) = \sqrt{\frac{\log(2)}{\pi \alpha_D^i}}\exp\left( \frac{(f - f^i_c)^2log(2)}{c\alpha_D^i} \right).
\end{equation}

The important thing here is which line shape to choose in the case of paths that penetrate vast vertical distances in the atmosphere. As we move from lower altitudes to higher altitudes, we need to decide which line shape is the appropriate one for calculating the line shape. In the transition zone, i.e., where Doppler and Lorentz half widths are comparable, one should utilize Voigt line shape \cite{Huang2004}. This can be obtained as a convolution of the Lorentz and Doppler line shapes by
\begin{equation}
F^i_V(f) = \frac{1}{\sqrt{\pi}\alpha_D}\frac{1}{\pi}\frac{\alpha^i_L}{\alpha_D}\int_{\pm\infty}\frac{\exp(-t^2)}{[\frac{f-f_c}{\alpha_D}-t]^2 + (\frac{\alpha^i_L}{\alpha_D})}dt.
\end{equation}
In our calculation, we utilize Voigt line shape for those lines that have Doppler and Lorentz half widths within five times of each other. Otherwise, the dominant line shape is utilized, i.e., the one that produces larger line width.


The line intensity $S_0^i$ can be obtained from HITRAN database for reference temperature $T_0$ = 296 K, but it has to be scaled for the other temperatures with \cite{HITRAN96}
\begin{equation}
S^i(T) = S_0^i\frac{Q(T_0)}{Q(T)}\frac{e^{(-\frac{hcE_L^i}{k_BT})}}{e^{(-\frac{hcE_L^i}{k_BT_0})}} \left(\frac{1-e^{(-\frac{hf_c^i}{k_BT})}}{1-e^{(-\frac{hf_c^i}{k_BT_0})}}\right)
\end{equation}
where $E_L^i$ is the lower state energy of the transition of absorbing species $i$. Notice that we utilize HITRAN database where the lower state energies have been given in the units 1/cm. Thus, the multiplication of $E_L^i$ with $hc$. The partition function $Q(T)$ and its definitions can be found in \cite[Appendix A]{HITRAN96}.

Putting the absorption loss model together, the altitude dependent absorption coefficient becomes
\begin{equation}
  \kappa_a^i(f,r_\text{atm},p,T)=\frac{p\mu_i N_A}{RT} S^i(T)F^i(f,p,T),
\end{equation}
where distance $r_\text{atm}$ through the atmosphere is indirectly having impact on the parameters of the model depending on the altitude of the layer, but not to the actual distance through the layer which is handled by the integration over the total molecular loss in the atmosphere. The line width in the above is
\begin{align}
  F^i(f,p,T) = \left \{\begin{array}{lll}
      F^i_{VVH}(f,p,T), &\text{if} ~\alpha_L^i >> \alpha_D^i,\\
      F^i_{D}(f,T),  &\text{if} ~\alpha_L^i << \alpha_D^i,\\
      F^i_V(f,p,T), &\text{if} ~\alpha_L^i \approx \alpha_D^i.\\
    \end{array}
  \right.
\end{align}
Notice that the altitude defines the pressure, temperature, and other parameters utilized in the above equations. As a consequence, the line shape function requires a dynamic algorithm that defines the correct line shape for every possible set of pressure, temperature, and absorption line center frequency. This can easily be done in the line-by-line calculation by calculating and comparing the Doppler and Lorentz absorption line half-widths.

\subsection{Transmission Path Geometry}

\begin{figure*}[t!]
    \centering
    \includegraphics[width=6in]{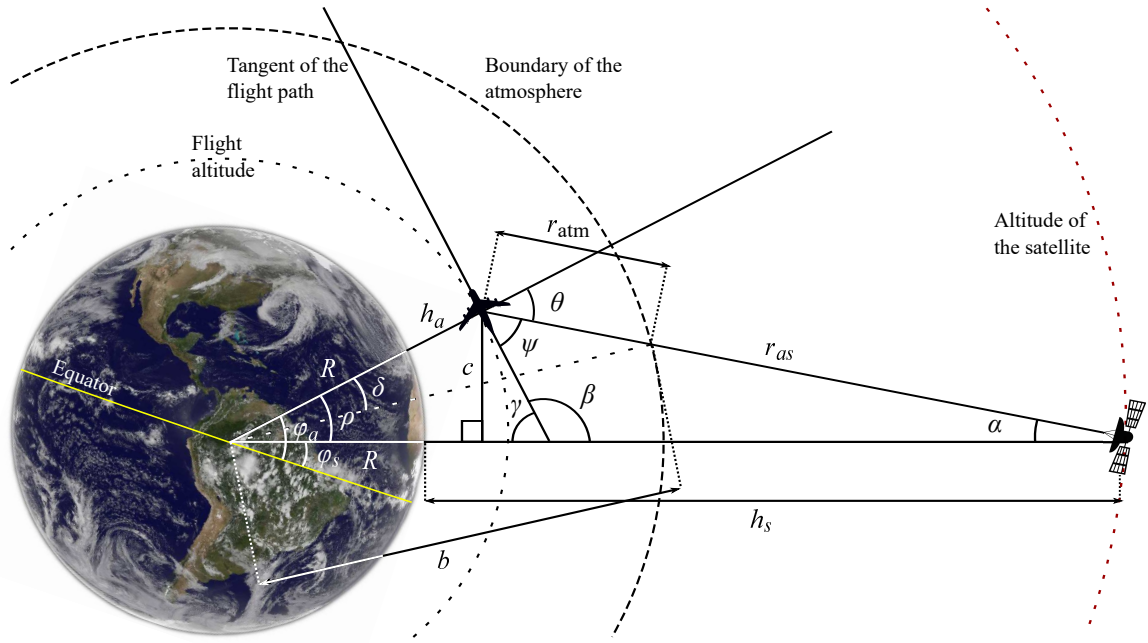}
    \caption{The system geometry for the angles and distances in the airplane to satellite system. Picture of Earth by NASA/NOAA's GOES Project \cite{NASApic}.}
    \label{fig:sysgeom1}
\end{figure*}

\begin{figure}[!h]
    \centering
    \includegraphics[width=3.2in]{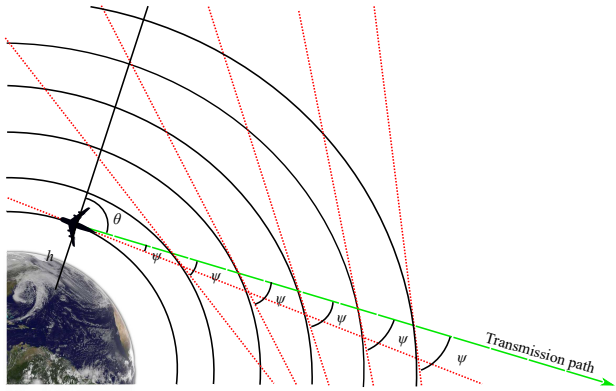}
    \caption{System geometry on the angles of the transmission path through the atmosphere. Picture of Earth by NASA/NOAA's GOES Project \cite{NASApic}.}
    \label{fig:sysgeom2}
    \vspace{-4mm}
\end{figure}

The general system geometry is shown in Fig.~\ref{fig:sysgeom1}. The usual way to model the atmosphere is to assume plane parallel atmosphere. By assuming flat surface, the geometry of the system is significantly simplified. However, when talking about applications where the entire atmosphere is penetrated, the plane parallel assumption overestimates the thickness of the atmosphere. The impact is the higher the lower is the angle to the surface. In order to rectify the problems with the plane parallel model, in this work we model realistic curving atmosphere. In such setting, the problem becomes with ever-changing angle to the zenith (and to the elevation angle $\psi$) as shown in Fig. \ref{fig:sysgeom2}. This problem can be modeled geometrically by utilizing the geometry shown in Fig. \ref{fig:sysgeom1}. In this figure, $R$ is the Earth's radius, $h_a$ is the altitude of the airplane, $b$ is the Earth's radius plus the total thickness of the atmosphere, $h_s$ is the altitude of the satellite, $\phi_a$ is the latitude of the airplane, $\phi_s$ is the latitude of the satellite, $r_{as}$ is the distance from the airplane to satellite, and $r_\text{atm}$ is the most interesting parameter, i.e., the distance through the atmosphere from the airplane to the boundary of the atmosphere. This latter distance is important because of the molecular absorption that depends on the altitude and ultimately the layer thicknesses given certain resolution, and according to the above absorption and atmospheric models.


The main difference of the curving atmosphere to the plane parallel atmosphere comes from the dynamically changing atmospheric parameters depending on the altitude. This is shown in Fig. \ref{fig:sysgeom2}. Distance from the airplane to satellite is given by the law of cosines
\begin{equation}
r_{as} = \sqrt{(R+h_a)^2 + (R+h_s)^2 - 2(R+h_a)(R+h_s)cos(\rho)},
\end{equation}
where $R$ is the radius of Earth, $h_a$ is the altitude of the airplane, $h_s$ is the altitude of the satellite, and $\rho$ is the angle between the airplane and satellite looked from the center of the Earth. This angle is obtained as subtraction of the elevation angles, i.e., $\rho = |\phi_s - \phi_a|$, where $\phi_s$ is the elevation angle of the satellite and $\phi_a$ is the same for the airplane. See details on these variables in Fig. \ref{fig:sysgeom1}. The distance from airplane to imaginary line from Earth's core to satellite is calculated as
\begin{equation}
c = (R+h_a) \sin(\rho).
\end{equation}
then the angle between the satellite and the airplane, looked from the satellite, is obtained from the two as
\begin{equation}
\alpha = \sin^{-1}\left( \frac{c}{r_{as}} \right).
\end{equation}
The elevation angle of transmissions $\psi$, looked from the airplane to satellite, is obtained simply as
\begin{equation}
\psi = 180 - \beta - \alpha = 180 - (180 - \gamma) - \alpha = \gamma - \alpha,
\end{equation}
where $\gamma = 90 - \rho$. Now we have everything we need in order to calculate the distance from airplane through the atmosphere $r_\text{atm}$. This can be calculated from the triangle suspended by $R+h_a$, distance from the core of the Earth to boundary of the atmosphere $b$, and $r_\text{atm}$. We can solve $r_\text{atm}$ with the law of cosines and the quadratic formula as
\begin{equation}
\begin{aligned}
r_\text{atm} &= (R+h_a) \cos(\psi+90) \\&+ \frac{1}{2}\sqrt{(-2(R+h_a) \cos(\psi+90))^2-4((R+h_a)^2-b^2)}.
\end{aligned}
\end{equation}
Finally, in order to fit this new approximation for the length of the atmosphere to the transmittance, we need to take a derivative of it. The reason is that the absorption coefficient changes as a function of the altitude, but does not depend on the total distance. Therefore, we need to manipulate the space over which the total absorption is calculated. That is, we multiply the absorption coefficient by $d/dr(r_\text{atm})$ instead of $\sec(\theta)$ as in the case of plane parallel atmosphere. Then,
\begin{equation}
\label{equation:tau_curve}
\tau(f,r) = e^{-\int\limits_{r_1}^{r_2}\kappa_{a}(f,r)d/dr(r_\text{atm})dr} = e^{-\int\limits_{r_1}^{r_2}\kappa_{a}(f,r_\text{atm})dr_\text{atm}}.
\end{equation}

\begin{figure}[t!]
    \centering
    \includegraphics[width=3.2in]{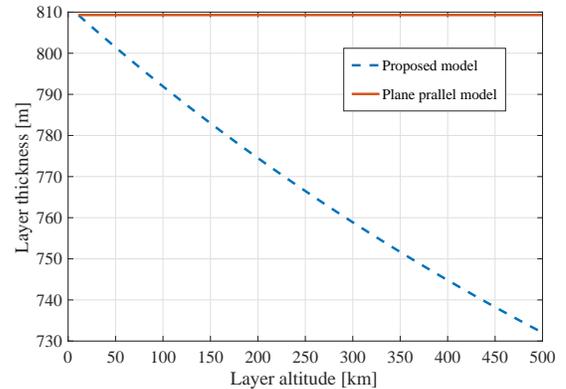}
    \caption{Comparison of the layer thickness of the plane parallel atmosphere and a round atmosphere as a function of altitude.}
    \label{fig:layers}
    \vspace{-5mm}
\end{figure}

\begin{figure}[t!]
    \centering
    \includegraphics[width=3.2in]{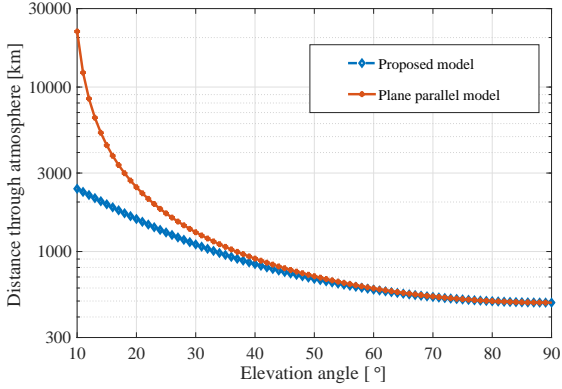}
    \caption{A comparison of the total distance through the atmosphere according to the plane parallel assumption and the model herein as a function of the elevation angle.}
    \label{fig:paths}
\end{figure}

The difference between the plane parallel assumption and the proposed model is shown in Figs. \ref{fig:layers} and \ref{fig:paths}. In computer simulations/models, we are forced to discretize the parameters due to utilization of the data bases to obtain the parameters and since the computer inherently handles only discrete data. Then we also have to discretize the atmosphere into equally spaced layers. In the plane parallel atmopshere, the distance through each layer is same at all altitudes due to the simple geometry. In the geometry given above, the dicretization leads into non-equal propagation distance through the layer depending on the altitude and the angle of penetration due to curvature of the atmosphere. It can be seen in Figs. \ref{fig:layers} and \ref{fig:paths} that the vertical path through the atmosphere is exactly the same for both model as they should be. Notice that the atmosphere has been calculated up to 500 km altitude where it is already very thin and does not contribute much on the propagation anymore. Fig. \ref{fig:layers} shows the propagation distance through a single 500 m layer (500 m resolution accuracy) at 38.2 degree angle from the ground. As stated above, the layer altitude has impact on the dynamic distance through the atmosphere. Thus, the plane parallel assumption tends over estimate the layer thickness. The problem is the worse the lower is the angle through the atmosphere. This is well visible in Fig. \ref{fig:paths}, which shows the total distance through the atmosphere. As the molecular absorption is dependent on the propagation through molecular medium, the plane parallel assumption may greatly overestimate the total absorption loss. The next section looks into derivation of the absorption coefficient $\kappa_a(f,r)$ as the geometry of the propagation is one important issue, and the second important aspect are the parameters fed into the model.

\subsection{Total Path Loss}


By combining \eqref{equation:FSPL} with \eqref{equation:tau_curve}, the total path loss can be written as
\begin{equation}
    PL(f,r) = \frac{(4\pi r f)^2\delta_\text{Rain}(r_r)\delta_\text{Cloud}(r_c)}{c^2 G_{Tx}(\theta_{Tx})G_{Rx}(\theta_{Rx})}e^{\int\limits_{r_1}^{r_2}\kappa_{a}(f,r_\text{atm})dr_\text{atm}},
    \label{eq:pl}
\end{equation}
where $G_{Tx}(\theta_{Tx})$ and $G_{Rx}(\theta_{Rx})$ are the Tx and Rx antenna gains towards directions $\theta_{Tx}$ and $\theta_{Rx}$, respectively, and $\delta_\text{Rain}(r_r)$ and $\delta_\text{Cloud}(r_c)$ are the channel gains due to rain and cloud losses, where $r_r$ is the distance through the rain and $r_c$ is the distance through the cloud/fog.


\section{Received Signal Power}
\label{sec:snr}


In order to calculate the SNR and subsequently the channel capacity of the links under study, this section studies the properties of the received signal including the noise. 

\subsection{Total Received Signal Power}


The received signal power spectral density $Y$ at the receiver is given by
\begin{equation}
Y\left(f,r\right) = X\left(f\right) PL(f,r) + N\left(f,r\right),
\end{equation}
where $X$ is the transmitted signal power spectral density with total power $P_{Tx}=\int_W X\left(f\right) df$ and $W$ is the signal bandwidth, $PL$ is the total path loss given by~\eqref{eq:pl}, and $N$ refers to the total noise power spectral density. As we discuss next, the calculation of the noise at THz frequencies is a non-trivial process.  

\subsection{Noise}

The last piece of the puzzle before calculating the SNR is the noise. The noise is an interesting problem in THz band since the molecular absorption process introduces additional antenna noise due to atmosphere acting as black body radiator, i.e., the atmosphere emits radiation according to the Planck's law. At regular atmospheric temperatures, the atmospheric black body radiation begins to contribute significantly around  sub-THz frequencies. This can be modelled via antenna brightness temperature. There are two options: at lower frequencies, where $hf \ll k_BT$, the Rayleigh-Jeans law can be utilized \cite{Wilson2009}. At higher frequencies, this model suffers from so called ultraviolet catastrophe, i.e., it gives infinite overall black body energy due to ever-increasing energy towards higher frequencies. The second option is to use the Planck's law which is accurate everywhere. However, in the THz band, both models are rather accurate and one may want to choose the Rayleigh-Jeans law due to easier implementation. The brightness temperature according to the Rayleigh-Jeans law is \cite{Wilson2009}
\begin{equation}
T_b^R = T(1-e^{-\kappa_a(f) r}) = T(d)(1-e^{-\int\limits_{r_1}^{r_2}\kappa_{a}(f,r_\text{atm},p,T)dr_\text{atm}}),
\end{equation}
where $T$ is the temperature, or temperature profile of the atmosphere and the last term comes from the derived channel model and $1-e^{-\kappa_a(f) r}$ gives the emissivity of the atmosphere. The more general model for the brightness temperature can be derived from the Planck law \cite{Wilson2009}
\begin{equation}
B_f^P(f,T) = \frac{2hf^3}{c^2(e^\frac{hf}{k_B T}-1)}.
\end{equation}
Then the brightness temperature can be calculated as
\begin{equation}
T_b^P = \frac{hf}{k_B}\text{ln}\left(1 + \frac{e^{\frac{hf}{k_BT(d)}}-1}{1-e^{-\int\limits_{r_1}^{r_2}\kappa_{a}(f,r_\text{atm},p,T)dr_\text{atm}}}\right)^{-1}.
\end{equation}
We utilize Planck's law in this paper due to it's more general applicability. However, as mentioned above, a simpler approximation would be given by the Rayleigh-Jeans law that would be well applicable at frequencies below one terahertz.

Besides the atmospheric noise, the thermal noise at the receiver needs to be taken into account. The thermal noise begins to decrease at the THz frequencies due to quantum effects at very low temperatures, or very high frequencies, i.e., when $hf \gg k_bT$. At this region, the probabilities of the higher energy states of the molecules/atoms become smaller and smaller, making the thermal noise smaller and smaller with respect to the one predicted by simple $k_BT$ \cite{Nyquist1928}. Taking into account the total noise power degradation, the thermal noise power density with receiver noise figure $NF$ (in dB) becomes
\begin{equation}
\begin{aligned}
N_T(f,NF) &= k_BT\eta(f)10^{NF/10} \\&= k_BT\frac{hf/k_BT}{\exp(hv/k_BT)-1}10^{NF/10} \\&= \frac{hf}{\exp(hv/k_BT)-1}10^{NF/10},
\end{aligned}
\end{equation}
where function $\eta(f)$ takes into account the noise power reduction. However, $\eta(f)$ does not play crucial role in the THz communications, since the transition when quantum of energy exceeds per Hertz thermal noise power occurs at 6.168 THz at 296 K temperature, i.e., when $k_BT = hv$. Therefore, the noise reduction is modest, up to few dBs at 10 THz frequency. Considering the fact that the THz band systems are envisioned to be extremely wide band systems, there is far more noise than in the conventional systems, mainly contributed by the large thermal energy of the wide band systems.

Finally, the total noise power spectral density at the receiver becomes:
\begin{equation}
N\left(f\right) = B_f(f,T) + N_T(f,NF).
\end{equation}
The total noise power can be then obtained by integrating $N$ over the receiver bandwidth, $W$ in our model.


\subsection{The System SNR and Capacity}
Taking into account the noise and the path loss from above, the SNR becomes
\begin{equation}
\text{SNR}\left(f,r\right) = \frac{ X\left(f\right)PL\left(f,r\right)}{N \left(f\right)},
\end{equation}
and the corresponding Shannon capacity is
\begin{equation}
C\left(r\right) = \int_W \text{log}_2(1 + SNR\left(f,r\right)) df.
\end{equation}

\section{Numerical Results}
\label{section:res}
In this section, we evaluate the expected path losses and link performances based on the above channel models.


\subsection{Parameters for the Channel Model}

\begin{figure}[t!]
    \centering
    \includegraphics[width=3.2in]{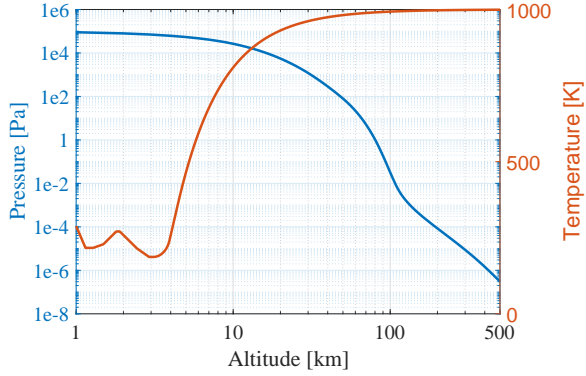}
    \caption{Atmospheric pressure and temperature as a function of altitude.}
    \label{fig:usstd1}
\end{figure}

\begin{figure}[t!]
    \centering
    \includegraphics[width=3.2in]{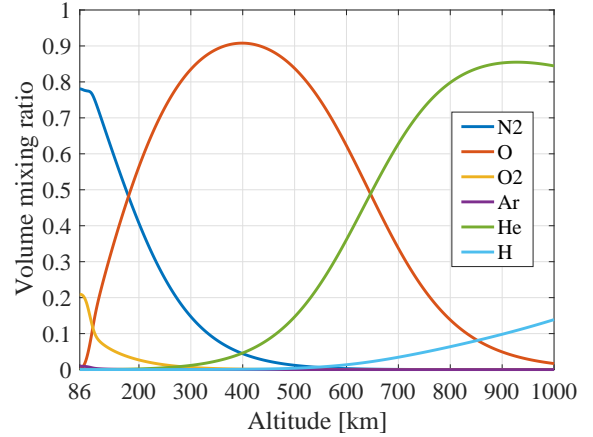}
    \caption{Volume mixing ratios of different molecules as a function of altitude.}
    \label{fig:usstd2}
\end{figure}

\begin{figure*}[t!]
    \centering
    \includegraphics[width=6.5in]{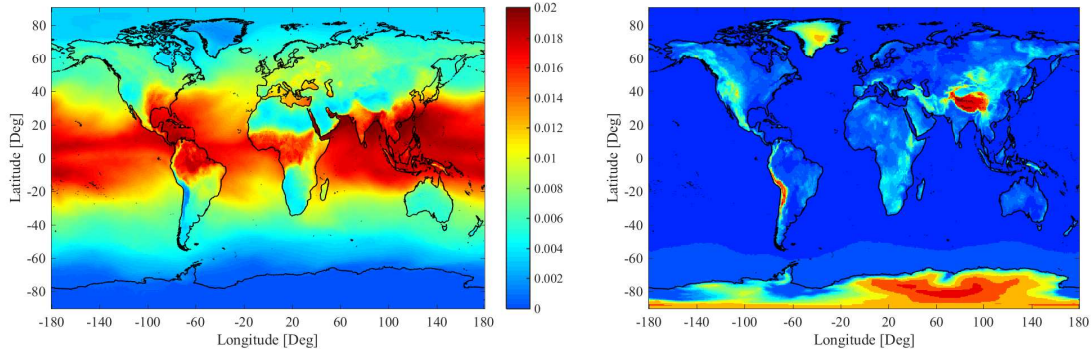}
    \caption{The surface volume mixing ratio of water vapor (left) and the surface height from the sea level (right).}
    \label{fig:GloHum}
\end{figure*}

One important factor in modeling the atmosphere is to model the altitude dependent parameters correctly. Those are mainly the temperature and pressure, and volume mixing ratios of different molecules. All these vary quite a bit as a function of altitude. For this work, we utilized the 1976 US Standard Atmosphere \cite{StdAtm1976}. The corresponding temperature and pressure as a function altitude is given in Fig. \ref{fig:usstd1}. The volume mixing ratios of common molecules is given in Fig. \ref{fig:usstd2}. Interestingly, the volume mixing ratios of the molecules remain roughly the same to the sea level up until 86 km altitude, after which they start to vary to each other. The pressure, temperature, volume mixing ratios, and ground level humidities are shown in Figs. \ref{fig:usstd1} to \ref{fig:GloHum}.


If one is interested modeling applications close to ground level, the global average humidities and the ground height has an impact on the molecular absorption to some extent. This mostly depends on the latitude, but also in the exact location in some cases. Fig. \ref{fig:GloHum} shows the average humidity around the world (left) and the ground height (right). The data for these were obtained from NASA/GMAO MERRA2 reanalysis data \cite{MERRA}. From \cite{MERRA}, one can find large datasets of atmospheric data, including global humidity data at different time scales and much more related to global weather and climate.

The rain and fog attenuations were shortly discussed above. In the real atmosphere, we do not always have clear weather and additional losses by clouds and rain impair the link performance. These are the mostly affecting low altitude communications, below few kilometer heights. As such, they have small impact on A2S and communications while at the typical flight heights. However, some losses by clouds can also exist at very high altitudes. For the take off and landing phases these may have an impact, and also in the G2A use case.

\begin{figure}[!h]
    \centering
    \includegraphics[width=3.2in]{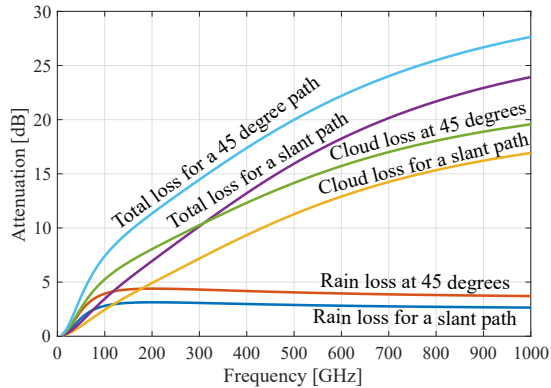}
    \caption{Rain and fog attenuation for slant paths as well as on 45 degree angled paths.}
    \label{fig:rainfog}
\end{figure}

In the below example shown in Fig. \ref{fig:rainfog}, we assume a typical nimbostratus cloud, which is roughly one kilometer thick and has a cloud bottom height of 0.7 km \cite{Slobin1982}. The cloud density is assumed to be 0.5 g/$\text{m}^3$, which is close to a typical density of a nimbostratus cloud \cite{Slobin1982}. The rain rate is assumed to be 5 mm/hr, which corresponds to a moderate rain. Figure \ref{fig:rainfog} shows the impact of the clouds and rain with the above assumptions for a slant path and path angled at 45 degrees. The angled path therefore increases the path lengths inside the rain and clouds by $\text{sec}(\theta  = 45^\circ)$, or $\sqrt{2}$. We can see that the losses vary from few decibels to few tens of decibels. Depending on the frequency and other losses, this may cause problems in signal reception and should be taken into account in the link budget considerations. However, as mentioned above, the rain and cloud/fog attenuations are mostly affecting low altitide communications, such as E2A communication scenario.
%


\subsection{Path Losses}
\label{subsec:results_a2s}

\begin{figure}[t!]
    \centering
    \includegraphics[width=3.2in]{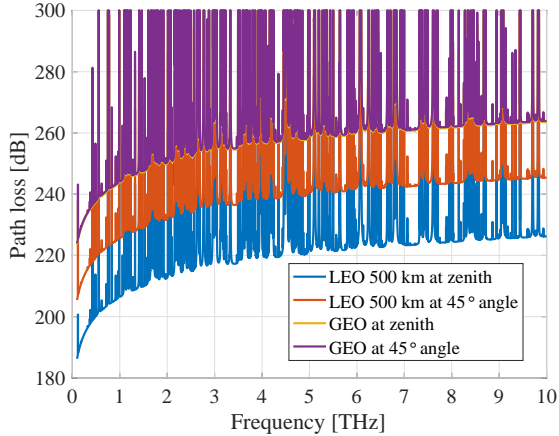}
    \vspace{-2mm}
    \caption{Path loss from airplane to satellite for LEO and GEO orbits as a function of frequency. The airplane altitude is 11 km.}
    \label{fig:AtoS}
    \vspace{-2mm}
\end{figure}

Figure \ref{fig:AtoS} shows the path loss from airplane to satellite for an airplane at 11 km altitude as a function of frequency. The path losses are shown for a geostationary Earth orbit (GEO, $\sim$36000 km) and for a low Earth orbit (LEO) at 500 km altitude. The path losses have been calculated for zenith path and for a 45$^\circ$ paths through the atmosphere. The LEO connection suffers more from the angled paths due to the path loss by absorption has more relatively higher impact on the shorter paths. The GEO path loss is dominated by the FSPL and the absorption does not have a that large impact on the total path loss. We can see that the path losses are considerable. Those range from about 190 to 260 dBs disregarding the absorption peaks that are much higher. This means that the required antenna gains need to be very high, but at the same time, the bandwidths need to be low enough to keep the noise level down and to avoid the absorption peaks. Considering fixed aperture antennas, the antenna diameters need to be in the order of 0.5 - 1 meters to provide enough gain in the ideal case.

\begin{figure}[t!]
    \centering
    \includegraphics[width=3.2in]{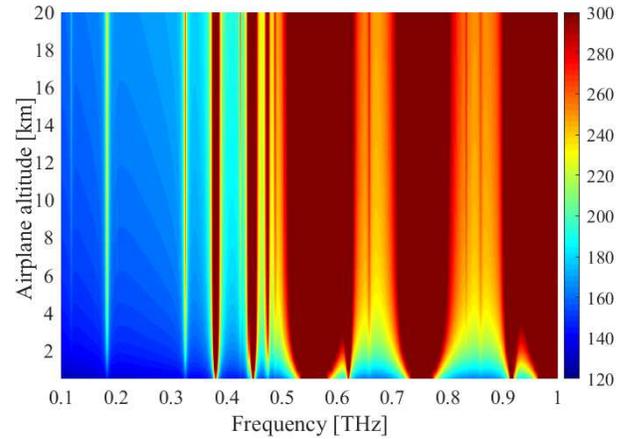}
    \caption{Path loss from Earth to airplane as a function of airplane altitude and frequency.}
    \label{fig:EtoA}
\end{figure}

\begin{figure}[t!]
    \centering
    \includegraphics[width=3.2in]{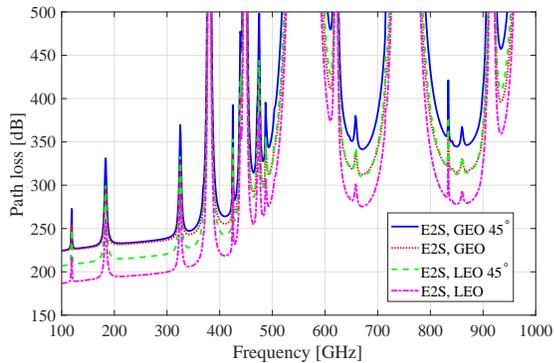}
    \caption{Path loss from Earth to satellite as a function of frequency.}
    \label{fig:EtoS}
\end{figure}

The path losses from Earth to airplane as a function of airplane altitude are shown in Figs. \ref{fig:EtoA} and \ref{fig:EtoS} for a frequency range from 100 GHz to 1000 GHz. The latter figure assumes LEO height of 500 km. Any frequencies above 1000 GHz are effectively blocked due to high absorption loss close to Earth. We can see that the below 380-450 GHz frequencies are utilizable for airplane communications from the ground. However, the losses remain high and partially so due to high FSPL at high frequencies. This also means a certain type of a bottle neck for feeding the satellites. The airplane to satellite links suffer less from the atmospheric loss and could theoretically utilize higher frequencies that than the Earth to airplane, or Earth to satellite links. This potentially opens a door for frequency division between ground to satellite and air to satellite links. However, this is also dependent on the capacity requirements, and hardware limitations. The latter is a major problem at the THz frequencies and potentially considerable losses have to be expected from inefficiency of them. In this paper we do not consider those, as the development of the THz band hardware is still a major research topic and relatively poorly understood from the viewpoint of realistic system and link performance. It should still be kept in mind that more losses are expected compared to pure atmospheric losses due to rain, clouds/fog, and other possible mechanism, such as the scattering loss.

\begin{figure}[t!]
    \centering
    \includegraphics[width=3.2in]{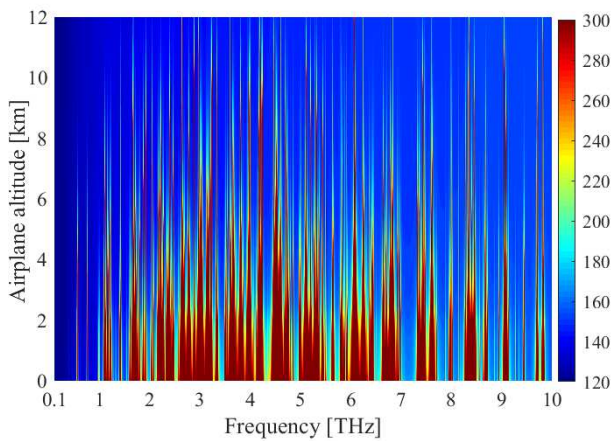}
    \vspace{-2mm}
    \caption{Path loss from airplane to airplane (common altitude) for a 100 meter link as a function of airplane altitude and frequency.}
    \label{fig:AtoA}
    \vspace{-2mm}
\end{figure}

\begin{figure}[t!]
    \centering
    \includegraphics[width=3.4in]{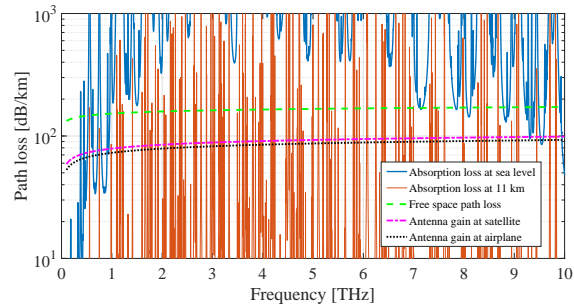}
    \caption{Comparison of path losses per kilometer for FSPL, molecular absorption loss at sea level and 11 km altitude, and antenna gains of 0.5 and 1 meter ideal dish antennas.}
    \label{fig:PLperkm}
\end{figure}

Airplane to airplane path loss for a 100 meter link is shown in Fig. \ref{fig:AtoA} as a function of altitude and frequency. As expected, the path loss decreases with altitude and makes the communications more simple. At 12 km height, the loss is mostly comprised of the FSPL disregarding some spikes in the spectrum. This is also shown in Fig. \ref{fig:PLperkm}, which shows the path losses per kilometer link on ground level and at 11 km height. As a reference, the antenna gains for 0.5 m diameter dish antenna ('airplane antenna') and 1 meter diamter dish antenna ('satellite antenna') are also shown. It should be noticed that the airplanes in general fly far away from each other and 100 meter distance would be considered very dangerous. The airplane to airplane links do not have many use cases in civil aircrafts, but would be potential for military applications and in drone-to-drone applications. Also, these figures are applicable to on Earth communications due to varying land height. On a general note on the high frequency links, the THz band offers a natural protection against third parties to listen to the data traffic as the FSPL and atmospheric losses kill the signals over long distances in the lower atmosphere as seen in Fig. \ref{fig:EtoA} and \ref{fig:AtoA}.

\subsection{SNR and Capacity}
\label{subsec:results_snr}

\begin{figure}[t!]
    \centering
    \includegraphics[width=3.2in]{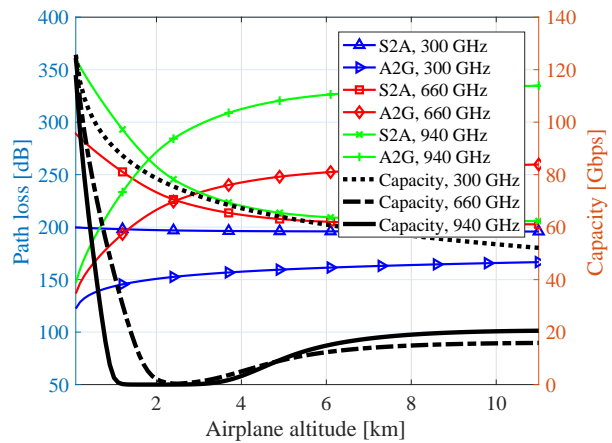}
    \caption{Comparison of A2G and A2S path losses at 300, 660, and 940 GHz center frequencies as a function of the airplane altitude as well as the capacity of the minimum loss path.}
    \label{fig:G2A2S}
\end{figure}

Based on the path losses, we can estimate the performance of the extremely long links. Figure \ref{fig:G2A2S} shows the path losses from A2G and S2A (or G2A and A2S) as a function of the airplane altitude. The satellite is on LEO at 500 km height. As it can be expected, there is a cross over altitude for the lower loss connection direction. This cross over point depends on the frequency, but suggests that the best way to provide THz connectivity would be to have both links available. However, during the mid-flight, the satellite link provides better connectivity. Considering that the use of electronic devices is not allowed during the take off and landing, the satellite link alone would be enough. It should also be noticed that the calculation herein does not take into account the movement of the airplane that would have more severe impact towards ground since the link distance is much lower that towards the satellites. The true path loss towards the ground would therefore also depend on the available ground stations and their locations. Figure \ref{fig:G2A2S} also provides the capacity figures for the calculated path losses assuming 1 mW transmit power, 5 GHz bandwidth, 10 dB noise figure, and 0.5 meter dish antenna at airplane and 1 meter dish antenna at satellite and ground. The capacities can reach up to 120 GHz at close proximity to Earth station. Mid-flight capacities ranging from 20--60 Gbps can be achieved.

\begin{figure*}[t!]
    \centering
    \includegraphics[width=6.4in]{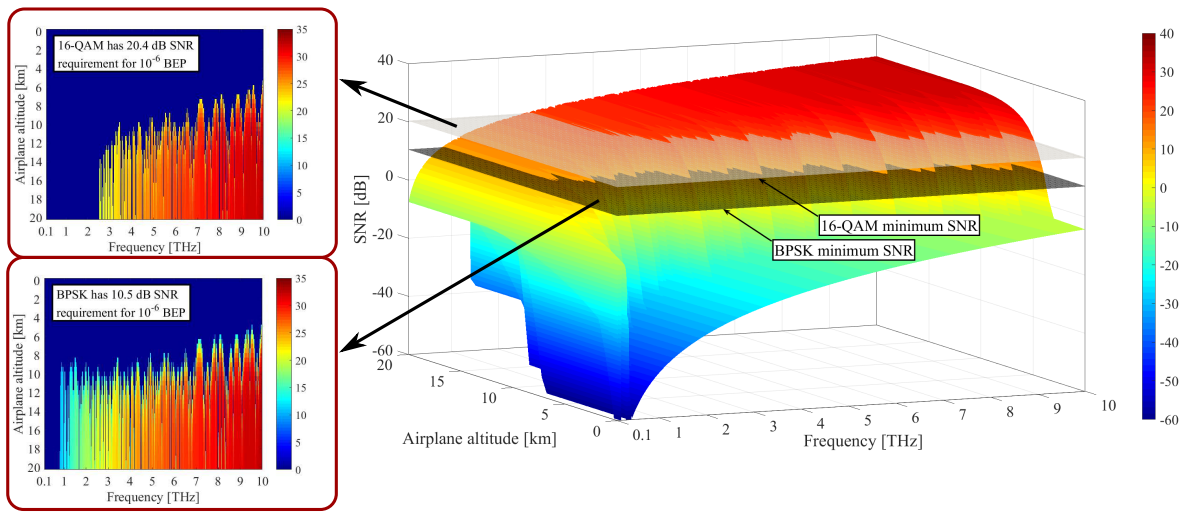}
    \vspace{-2mm}
    \caption{SNR from satellite to airplane as a function of airplane altitude and frequency for GEO orbit. Figure also shows boundaries for square gray coded 16-QAM and BPSK for target BEP probability of $10^{-6}$.}
    \label{fig:SNR_GEO}
    \vspace{-2mm}
\end{figure*}

The expected SNR as a function of altitude and frequency for A2S link to GEO is given in Fig. \ref{fig:SNR_GEO}. It is assumed that the airplane has a 0.5 meter dish antenna and the satellite has a 1 meter dish antenna. As a consequence of the fixed apertures, the effective antenna gain increases as a function frequency. This is seen as one of the potential source to compensate the high losses in the channel; It will be easier to make electrically large antennas at THz frequencies that significantly increase the antenna gains \cite{Hwu2013}. This translates into increasing SNR as a function frequency in Fig. \ref{fig:SNR_GEO}. The SNR is further increased by the altitude of the plane due to before mentioned lower absorption loss at high altitudes. Figure \ref{fig:SNR_GEO} also shows the SNR limits of some modulation methods for 10$^{-6}$ targer BEP values. Namely, for 16-QAM and BPSK. As expected of the long distance links in the THz band, the usable modulation orders remain modest to ensure good BEP performance. It should be noted that the BER values are pure physical layer BEPs and do not take into account coding. Thus, the actual BERs would be lower with lower overall throughput due to coding overhead.

Whereas the overall picture favor the higher frequencies because of theoretically higher achievable antenna gains, the reality is that the hardware imperfections would most likely decrease the performances as the frequency is increased. However, the results herein show great potential for the THz frequency band for long distance communications. There are many obstacles ahead to realize these types of systems. Most notably in the hardware side, but with time and leaps the THz communications has taken during the past few years, these frequencies will truly be conquered in the near future. This also then opens doors for wide range of multiple scales of communications on these frequencies. As an example, to provide high speed satellite communications for the airplanes.


\section{Conclusion}
\label{section:concl}
We presented a channel model for airplane communications in the THz band in this paper. The results show that the path loss is very high, as it can be expected from an extremely high distance links operating at very lossy THz frequencies. However, if the antenna apertures are high, potentially very large antenna gains can be achieved. Regardless of the very high path loss, airplane to satellite, and airplane to airplane communications are perfectly possible on rather wide spectrum. The Earth to satellite or airplane case is more demanding due to high absorption loss close to Earth. This may limit the Earth communications to below 380 GHz frequencies. In practice, the 300 GHz frequencies are in any case the first frequencies to be commercially utilized in the near future. Possible weather impacts will decrease the signal levels, and increasingly so at higher end of the THz band. The frequency regulation, on the other hand, will limit the full utilization of the THz frequencies in the space applications due to protection of the passive Earth observation services. Although it is often said that the THz frequencies are only for short distance communications, they also have great potential in long distance applications. This is especially the case in space where the atmosphere causes less channel losses.

%


\ifCLASSOPTIONcaptionsoff
  \newpage
\fi

\bibliographystyle{IEEEtran}
\bibliography{sat_thz_01.bbl}

\end{document}